\documentclass{iaus}
\usepackage{graphicx}

\title[JD 11.~~Masers in low-mass \\star formation regions] 
{Class I methanol masers in low-mass \\star formation regions.}

\author[S.V. Kalenskii, V.I. Slysh, L.E.B. Johansson, et al.]   
{S. V. Kalenskii$^1$,
\fbox{V. I. Slysh$^1$,}
\fbox{L. E. B. Johansson$^2$,}\\
P. Bergman$^2$,
S. Kurtz$^3$,
P. Hofner$^4$,
 \and C. M. Walmsley$^5$}

\affiliation{$^1$Astro Space Center, Lebedev Physical Institute, 
84/32 Profsoyuznaya st., Moscow, 117997, Russia \\ email: {\tt kalensky@asc.rssi.ru} \\[\affilskip]
$^2$Onsala Space Observatory, Chalmers University of Technology, 439 92 Onsala, Sweden \\email: {\tt pbergman@chalmers.se} \\[\affilskip]
$^3$Centro de Radioastronom\'\i a y Astrof\'\i sica,  Universidad Nacional Autonoma de M\'exico (Morelia, Michoac\'an, M\'exico) \\email: {\tt s.kurtz@crya.unam.mx} \\[\affilskip]
$^4$Physics Department, New Mexico Tech., 801 Leroy Pl., Socorro, NM 87801, and
National Radio Astronomy Observatory, Socorro, NM 87801, USA \\email: {\tt hofner\_p@yahoo.com} \\[\affilskip]
$^5$Osservatorio Astrofisico di Arcetri, Largo E. Fermi 5,1-50125 Firenze,
Italy \\email: {walmsley@arcetri.astro.it} }
\pubyear{2012}
\volume{287}  
\pagerange{xxx--xxx}
\jname{Cosmic Masers: from OH to H$_0$}
\editors{R.S. Booth, L. Humphreys \& W. Vlemmings, eds.}
\begin{document}

\maketitle

\begin{abstract} Four Class I maser sources were detected at 44, 84,
and 95 GHz toward chemically rich outflows in the regions of low-mass
star formation NGC~1333I4A, NGC~1333I2A, HH25, and L1157. One more maser 
was found at 36~GHz toward a similar outflow, NGC~2023.  Flux densities 
of the newly detected masers are no more than 18 Jy, being much lower than 
those of strong masers in regions of high-mass star formation. 
The brightness temperatures of the strongest peaks in NGC~1333I4A, HH25, 
and L1157 at 44 GHz are higher than 2000 K, whereas that of the peak 
in NGC~1333I2A is only 176 K. However, rotational diagram analysis showed 
that the latter source is also a maser. The main properties of the newly 
detected masers are
similar to those of Class I methanol masers in regions of massive
star formation. The former masers are likely to be an extension of
the latter maser population toward low luminosities of both the masers
and the corresponding YSOs. 
\end{abstract}

\firstsection 
\section{Introduction}
In spite of a number of observations and theoretical works, the nature 
of Class I methanol  masers is still unknown. This is partly because 
until recently these masers have been observed only in regions of massive 
star formation, which are typically distant (2--3 kpc from the Sun or 
farther) and highly obscured at optical and even NIR wavelengths. 
In addition, high mass stars usually form in clusters. These properties 
make it difficult to resolve maser spots and to associate masers with 
other objects in these regions. In contrast, regions of low-mass star 
formation are much more widespread and many of them are only
200--300 pc from the Sun; they are less heavily obscured than
regions of high-mass star formation, and there are many
isolated low-mass protostars. Therefore, the study of masers in
these regions might be more straightforward
compared to that of high-mass regions, and hence, the detection of 
Class I masers there might have a strong impact on maser exploration. 

Bearing this in mind, we undertook a search for Class I 
methanol masers in regions of low-mass star formation. Since the most common 
viewpoint is that these masers arise in postshock gas in the wings of 
bipolar outflows~(\cite[Plambeck \& Menten, 1990]{plammen}; 
\cite[Chen et al, 2009]{chen}) our source list was composed of these objects.
The naive expectation is to find methanol masers towards bright thermal 
sources of methanol; therefore the basis of our source list consists of 
``chemically rich outflows'', where methanol abundance is
significantly enhanced relative to that in quiescent gas. Because methanol 
enhancement has been detected in young, well-collimated outflows from 
Class 0 and I sources, we included several such objects in our list 
regardless of whether methanol enhancement had been previously found there.
A subsample of our list consisted of YSOs with known outflows and/or 
H$_2$O masers located in Bok globules. Like other 
objects from our list, these YSOs are typically isolated objects of low or
intermediate mass, located in nearby ($<$500~pc) small and relatively simple 
molecular clouds. In total, our source list consisted of 37 regions which harbor 
46 known outflows driven by Class 0 and I low-mass protostars, taken from the literature.
All of them were observed in the $7_0-6_1A^+$ transition at 44 GHz, where 
the strongest Class I masers have been found so far. In addition to the  $7_0-6_1A^+$ transition, most sources were observed 
in other Class I maser lines, namely, in the $4_{-1}-3_0E$ line at 36~GHz, 
in the $5_{-1}-4_0E$ line at 84~GHz, and in the $8_0-7_1A^+$ line at 95 GHz,
as well as the ``purely thermal'' $2_K-1_K$ lines at 96 GHz. 

\begin{figure}[t]
\label{04011flux}
\begin{center}
 \includegraphics[width=3.4in, angle=-90]{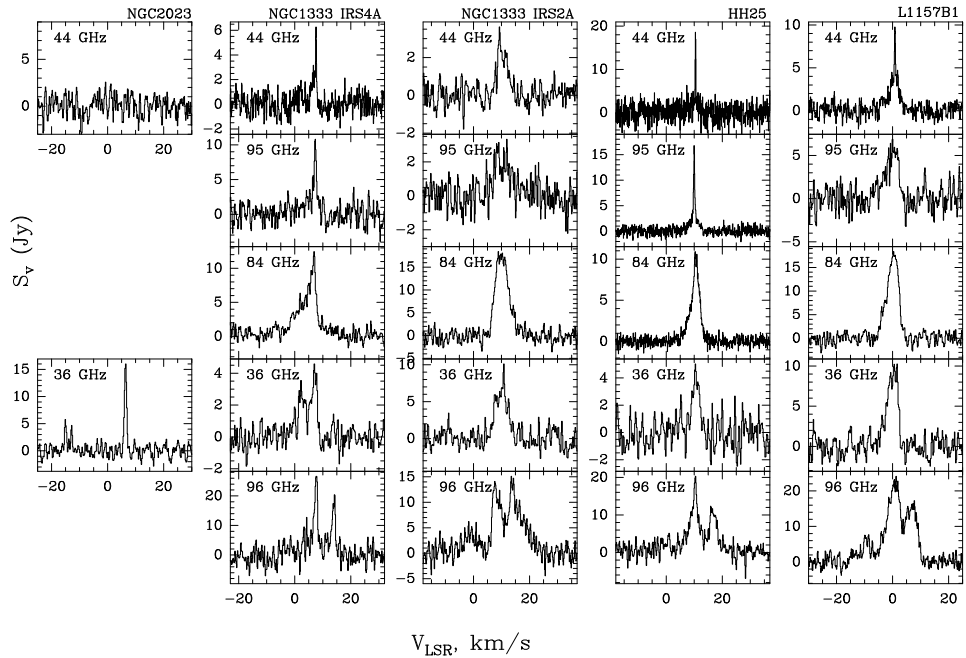} 
 \caption{Spectra of the newly detected masers.}
   \label{fig1}
\end{center}
\end{figure}
\section{Observations and results}
{\underline{\it Single-dish observations}.
The single-dish observations are described in detail by Kalenskii et al. (2006, 2010a).
They were carried out with the 20-m radio telescope 
of the Onsala Space Observatory (OSO) during several observing sessions in 2004--2011. 
As a result, we detected maser candidates at 44 GHz towards NGC~1333I2A, 
NGC~1333I4A, HH 25 and L1157. Toward NGC~1333I4A and HH~25, narrow
features were also found at 95 and 84 GHz. In addition, a narrow line
was detected at 36~GHz toward the blue lobe of an 
extremely high-velocity outflow in the vicinity of the bright reflection 
nebula NGC~2023. The source spectra are shown in Fig.~\ref{fig1}.

{\underline{\it VLA/EVLA observations}.
To check whether the newly detected sources are really masers we observed
them with the NRAO\footnote{The National Radio Astronomy Observatory is
operated by Associated Universities, Inc., under contract with the National
Science Foundation.} VLA/EVLA array in the D configuration, which provides 
an angular resolution about $1.''5$ at 44~GHz. L1157 was observed with 
the VLA on March 17, 2007; the other sources were observed with the EVLA 
on August 08, 2010. The data were reduced using the NRAO Astronomical 
Image Processing System (AIPS) package. The source parameters are presented
in Table~\ref{tab1}. 

\section{Are the newly detected sources really masers?}
The small sizes and high brightness temperatures at 44 GHz indicate that
the newly detected sources are masers. The exceptions are NGC~2023,
which was not found at 44 GHz, and NGC~1333I2A, with a line
brightness temperature of only 170 K~(Table~\ref{tab1}). 
The nature of the 36~GHz line in the blue lobe of the bipolar 
outflow in NGC~2023 is unclear. On the one hand, the line is fairly narrow, 
and offset measurements showed that the source is compact at least
with respect to the 105-arcsec Onsala beam. These properties 
suggest that 
the source is a maser. This assumption has further support in the fact 
that the line LSR velocity, $\approx 6.5$~km~s$^{-1}$, is less than 
the systemic velocity of about 10~km~s$^{-1}$. On the other hand, 
the line has no counterpart at 44~GHz, which is more typical for thermal
emission. Note, however, that there are known masers at 36~GHz without 44-GHz
counterparts; in particular, no 44 GHz emission was found at the velocity
of a fairly strong 36-GHz maser detected $3'$ north of DR21(OH) 
by~\cite{pratap}. Therefore, we tentatively conclude that the narrow line
in NGC~2023 is a maser.

The fairly low brightness temperature and finite sizes~(Table~\ref{tab1}) 
of NGC 1333I2A (M1 and M2) suggest that they are thermal sources. However,
a rotational diagram analysis (Kalenskii et al., in prep.) shows that
they are low-gain masers or a cluster of weak masers.

\begin{table}
  \begin{center}
  \caption{Parameters of maser sources determined by the VLA observations.}
  \label{tab1}
 {\scriptsize
  \begin{tabular}{|l|c|c|c|c|c|c|}\hline 
{\bf Source}  & {\bf R.A.} & {\bf Dec.} & {\bf Major}&{\bf Minor}&{\bf T$_{\bf BR}$}&{\bf V$_{\bf lsr}$} \\ 
              &{\bf(J2000)}&{\bf(J2000)}& {\bf axis ($\bf ''$)} & {\bf axis ($\bf ''$)}& {\bf (K)}        &{\bf (km s$^{\bf -1}$)}\\ 
\hline
NGC 1333I2A M1&03 29 00.802&31 14 21.32 & 2.2        & 1.4       & 170     & 10.9 \\
NGC 1333I2A M2&03 29 01.422&31 14 18.80 & 2.7        & 2.0       &  35     & 8.8 \\
NGC 1333I4A   &03 29 10.829&31 13 18.68 & 1.0        & 0.4       & 2400    & 6.9 \\
HH25 M1       &05 46 08.071&--00 14 05.66& 2.7       & 0.0       &  $\inf$ & 9.6 \\
HH25 M2       &05 46 07.967&--00 14 02.42& 0.7       & 0.0       &  $\inf$ & 9.6 \\
L1157 M1      &20 39 10.033&68 01 42.20 & 0.4        & 0.2       & 53000   & 0.8 \\
L1157 M2      &20 39 09.465&68 01 15.59 & 1.9        & 0.7       & 470     & 1.7 \\
 \hline

  \end{tabular}
  }
 \end{center}
\vspace{1mm}
\end{table}
 
\section{Properties of the new masers}
{\underline{\it Association with chemically rich outflows}}. 
New masers were found towards the lobes of outflows in 
NGC~1333I4A, NGC~1333I2A, NGC~2023, HH25, and L1157. These
outflows are known to be chemically rich outflows 
with enhanced methanol abundances.

Comparison of the VLA maps with high-resolution maps of thermal methanol
and other molecules shows that the masers coincide with chemically
rich gas clumps, where the abundances of methanol and other molecules
are enhanced (e.g., \cite{gibb98}; \cite{bach98}; \cite{ben07}). In L1157, 
the masers are located in gas clumps, which, according to chemical
modeling of~\cite{viti}, probably pre-existed the outflow.

{\underline{\it LSR velocities and intensities}}. 
Comparison of the maser LSR velocities with those of thermal methanol
lines, observed in the same directions, show that these velocities coincide 
within 0.5~km/s. This coincidence occurs even when the LSR velocities 
of some other molecular lines toward the maser positions are significantly
different. The LSR velocities of both maser and thermal methanol lines 
are usually close to the systemic velocities. 
An exception is the 36-GHz maser in the EHV outflow NGC~2023. 
Its radial velocity is less than the systemic velocity by 
about 3.5~km~$^{-1}$. Note that Voronkov (this volume) has detected 
a high-velocity Class I maser just at 36 GHz.

{\underline{\it Maser intensities}}. 
The new masers are weaker than the bright masers typical in regions of massive 
star formation. However, they obey the same relationship between the maser 
and YSO luminosities as reported by \cite{bae11} for
masers in regions of high- and intermediate-mass star formation,
thus extending this relationship toward low luminosities~(Kalenskii et al., in prep). 

{\underline{\it Variability}}. Several sessions of repeated 
observations of NGC 1333I4A, HH25, and L1157 at 44~GHz were performed 
in 2008--2011. No notable variations were found. Slight changes 
in line intensities can be attributed to poor signal-to-noise ratios
and calibration uncertainties. However, further monitoring of these 
sources is desirable in order to search for flares similar to that
which occurred in DR~21(OH).

To summarize, the main properties of the newly detected masers are
similar to those of Class I methanol masers in regions of massive
star formation. The former masers are likely to be an extension of
the latter maser population toward lower luminosities of both the masers
and the corresponding YSOs. 

\section{Maser models}
The fact that the maser LSR velocities coincide with the systemic
velocities allows us to conclude that the masers appear in dense clumps
of gas, probably pre-existing the outflows. However, the exact nature
of the masers remains unknown. \cite{sam98} suggested 
that compact maser spots arise in
extended, turbulent clumps because in a turbulent
velocity field the coherence lengths along some directions are larger
than the mean coherence length, resulting in a random increase of the
optical depth absolute values along certain sight lines in a clump.
According to ~\cite{kalen-b} such a model can easily explain the observed
brightness of the maser lines, but within the framework of this model
it is difficult to explain why {\em single peaks} dominate
the maser emission in the L1157 clumps. However, natural additional
assumptions, such as the existence of shocks or centrally condensed
clumps, makes it possible to explain the observational data.

An examination of the maser spectra in L1157 may lead to another
interpretation of our results. Both 44 GHz masers detected in this source
have double line profiles. It is known that a double thermal line with
a ``blue asymmetry'' may be a signature of collapse~\cite{zhou}. Contrary 
to this, the masers in L1157 exhibit a ``red asymmetry''. However, 
just such an asymmetry is what one would expect for Class I masers arising
in a collapsing clump.
This model is discussed in more detail by~\cite{kalen-b}. Note that this
model, if correct, is specific for the masers in L1157; no other
maser in our sample exhibits a double line profile. 

The work was financially supported by RFBR (grants No. 04-02-17547, 
07-02-00248, and 10-02-00147-a), and  Federal National Scientific 
and Educational Program (project number 16.740.11.0155).
P.H. acknowledges partial support from NSF grant AST 0908901. 
S.Kurtz acknowledges support from UNAM DGAPA grant IN101310.
The Onsala Space Observatory is the Swedish National
Facility for Radio Astronomy and is operated by Chalmers
University of Technology, G\"{o}teborg, Sweden, with financial
support from the Swedish Research Council and the Swedish Board
for Technical Development.



\end{document}